\documentclass[journal ]{IEEEtran}
\usepackage[T1]{fontenc}
\usepackage{color}
\usepackage{amsmath}
\usepackage{url}
\usepackage{array}
\usepackage{fancyhdr}
\usepackage{latexsym}
\usepackage{diagbox}
\usepackage{epsfig}
\usepackage{cite}
\usepackage{amssymb}
\usepackage{amsfonts}
\usepackage{amsxtra}
\usepackage{xspace}
\usepackage{makeidx}
\usepackage{graphics}
\usepackage{theorem}
\usepackage{algorithm,algpseudocode}
\usepackage{multirow}
\usepackage{epstopdf}
\usepackage{color}

\usepackage{bm}
\usepackage{subfigure} 

\begin{document}
	
	\title{Deep Learning based Denoise Network for CSI Feedback in FDD Massive MIMO Systems}
	
	\author{Hongyuan Ye, Feifei Gao, Jing Qian, Hao Wang, and Geoffrey Ye Li
%
%
   }
	
	\maketitle
	
	\begin{abstract}

Channel state information (CSI) feedback is critical for frequency division duplex (FDD) massive multi-input multi-output (MIMO) systems. Most conventional algorithms are based on compressive sensing (CS) and are highly dependent on the level of channel sparsity. To address the issue, a recent approach adopts deep learning (DL) to compress CSI into a codeword with low dimensionality, which has shown much better performance than the CS algorithms when feedback link is perfect. In practical scenario, however, there exists various interference and non-linear effect. In this article, we design a DL-based denoise network, called DNNet, to improve the performance of channel feedback. Numerical results show that the DL-based feedback algorithm with the proposed DNNet has superior performance over the existing algorithms, especially at low signal-to-noise ratio (SNR).
	\end{abstract}
	
	
	\IEEEpeerreviewmaketitle

	\section{Introduction}\label{sec:introduction}
	
Massive multi-input multi-output (MIMO) is widely recognized as the key technology for next-generation mobile communication systems since it can provide high transmission rate and high spectrum efficiency \cite{larsson2014massive,molisch2017hybrid,wang2018spatial-}. Despite its many advantages, the transmission quality of massive MIMO highly depends on the availability of channel state information (CSI). In a frequency division duplex (FDD) system, the base station (BS) cannot directly estimate the CSI of the downlink channel and such information is normally obtained through feedback channel. However, the amount of the feedback increases with the number of antennas, which leads to a huge overhead for massive MIMO systems.

Conventionally, compressive sensing (CS) is used to reduce feedback overheads by exploring the correlation among CSI \cite{huang2016rate}. However, CS-based algorithms may fail if the transmission environment is not strictly sparse. Recently, deep learning has been introduced to communications to improve its performance or to address some issues hard to be settled by traditional algorithms\cite{qin2019deep,aceto2019mimetic,aceto2019mobile,yang2019deep}. In this context, several deep learning (DL) based approaches \cite{wen2018deep,liu2019exploiting,guo2020convolutional} were proposed to mitigate the dependence on \textit{strict sparsity}. In \cite{wen2018deep}, an autoencoder model, named CsiNet, was developed for channel compression and reconstruction, where the estimated downlink CSI is first compressed into a codeword with low dimensionality at user equipment (UE), and is then recovered from the feedback codeword at the BS. CsiNet can achieve better feedback performance than the CS-based algorithms. In \cite{liu2019exploiting}, the available uplink CSI is exploited to recover downlink CSI to reduce CSI feedback overhead. In \cite{guo2020convolutional}, a convolutional neural network (CNN) based algorithm is designed to achieve multi-rate CSI feedback.

However, most of the existing algorithms ignore the influence of various interference and non-linear effects in practical feedback channel \cite{martins2008coding}. To the best of the authors' knowledge, the works that consider feedback error are \cite{jang2019deep,guo2020convolutional}. In this article, we design a separate DL-based denoise network, called DNNet, to overcome the the influence of feedback noise. The contributions are listed as follows:
    \begin{itemize}
    \item \textit{New denoise network design:} DnCNN \cite{Zhang2016Beyond} is originally used for image denoising. Since it cannot be directly applied to wireless communications, we modify the original structure for codeword denoising.
    \item \textit{New joint training mechanism:} We design a mechanism that jointly trains the existing DL-based channel feedback algorithm and the proposed DNNet, which has better performance than training them separately.
    \end{itemize}	

The rest of the paper is organized as follows. Section II introduces the system model, summarizes few existing DL-based CSI feedback algorithms, and then analyses their performance under non-ideal feedback. Section III presents the structure and training mechanism of the proposed DNNet for attenuating the effects of non-ideal feedback. Numerical results are demonstrated in Section IV and conclusions are provided in Section V.


	\section{Problem Formulation}\label{sec:systemmodel}
In this section, we first describe the model of the CSI feedback problem and then introduce the basic principle of the existing DL-based CSI feedback algorithms. Afterwards, we illustrate the problems that DL-based algorithm encounter under various interference and non-linear effects.
\subsection{System Model}
Consider an FDD massive MIMO system, where the BS has $N_{\rm t} $  antennas and UE has a single antenna. Orthogonal frequency division multiplexing (OFDM) with $N_{\rm c}$ subcarriers is deployed to convert the frequency-selective fading channel into multiple flat fading channels. Denote the CSI at the $k$th subcarrier as $\mathbf{h}_k \in \mathbb{C}^{N_{t} \times 1}$. Then the equivalent channel matrix can be expressed as
\begin{equation}
\label{E2}
\tilde{\mathbf{H}}=\left[\begin{array}{cccc}{\mathbf{h}_{1}^{H}} & {\mathbf{0}^H} & {\cdots} & {\mathbf{0}^H} \\
                                                        {\mathbf{0}^H} & {\mathbf{h}_{2}^{H}} & {\cdots} & {\mathbf{0}^H} \\
                                                        {\vdots} & {\vdots} & {\ddots} & {\vdots} \\
                                                        {\mathbf{0}^H} & {\mathbf{0}^H} & {\cdots} & {\mathbf{h}_{N_c}^{H}}\end{array}\right]_{N_{c} \times N_{c}N_{t}}.
\end{equation}
Let $\mathbf{p}_k \in \mathbb{C}^{N_{t} \times 1}$ represent the beamforming vector at the $k$th subcarrier, and then the equivalent beamforming matrix can be written as

\begin{equation}
\label{E3}
\mathbf{P}=\left[\begin{array}{cccc}{\mathbf{p}_{1}} & {\mathbf{0}} & {\cdots} & {\mathbf{0}} \\
                                                        {\mathbf{0}} & {\mathbf{p}_{2}} & {\cdots} & {\mathbf{0}} \\
                                                        {\vdots} & {\vdots} & {\ddots} & {\vdots} \\
                                                        {\mathbf{0}} & {\mathbf{0}} & {\cdots} & {\mathbf{p}_{N_c}}\end{array}\right]_{N_{c}N_{t} \times N_{c}}.
\end{equation}
The beamforming vector should meet the power constraints, i.e. , $\|\mathbf{p}_i\|_2^2 = P_t/N_c$, where $P_t$ denotes the total transmit power.
The downlink frequency domain signal vector received at UE can be written as
\begin{equation}
\label{E1}
\mathbf{y}=\tilde{\mathbf{H}} \mathbf{P} \mathbf{x}+\bm{\omega},
\end{equation}
where $\mathbf{x} \in \mathbb{C}^{N_{c} \times 1}$ and $\bm{\omega} \in \mathbb{C}^{N_{c} \times 1}$ represent the transmitted symbol vector and the noise vector, respectively.

For convenience, we rearrange the channel vector as
\begin{equation}
\label{E4}
{\mathbf{H}_{sf}}=\left[\begin{array}{cccc}{\mathbf{h}_{1}} & {\mathbf{h}_{2}} & {\cdots} & {\mathbf{h}_{N_c}}
                                                        \end{array}\right]^{H}_{N_{c} \times N_{t}},
\end{equation}	
where $\mathbf{H}_{sf}$ can be deemed as  the CSI in spatial-frequency domain. To extract the features of CSI, we take the 2D discrete Fourier transform (DFT) and transfer $\mathbf{H}_{sf}$ into the angle-delay domain as
\begin{equation}
\label{E5}
\mathbf{H}_{ad}=\mathbf{F}_{\mathrm{d}} {\mathbf{H}_{sf}} \mathbf{F}_{\mathrm{a}}^{H},
\end{equation}
where $\mathbf{F}_{\mathrm{a}}$ and $\mathbf{F}_{\mathrm{d}}$ are DFT matrices with dimensions $N_{c}\times N_{c}$ and $N_{t}\times N_{t}$, respectively. In the delay domain, CSI exhibits \textit{sparsity}, where $\mathbf{H}_{ad}$ has significant values only in the first $N_p$ rows since the time of arrival (TOA) between multipath belongs to a limited period \cite{wen2018deep}. Similar to \cite{wen2018deep,guo2020convolutional}, we choose the first $N_p$ rows of $\mathbf{H}_{ad}$ to form a new channel matrix $\mathbf{H}$ as
\begin{equation}
\label{E6}
    \mathbf{H} = [\mathbf{H}_{ad}]_{1:N_p,:},
\end{equation}
where $[\mathbf{H}_{ad}]_{1:N_p,:}$ denotes a sub-matrix that makes up of the first $N_p$ rows of $\mathbf{H}_{ad}$.
\subsection{DL-based Algorithm}
Recently, DL-based algorithms have been deployed in CSI feedback to reduce the transmission overheads \cite{wen2018deep,guo2020convolutional,liu2019exploiting}. The CSI at UE is first compressed into a codeword by an encoder,
\begin{equation}
\label{E7}
    \mathbf{s} = \bm f_{ en}(\mathbf{H}),
\end{equation}
and then the codeword is transmitted to the BS through a feedback channel. The CSI can be reconstructed by a decoder at the BS as
\begin{equation}
\label{E8}
    \hat{\mathbf{H}} = \bm f_{ de}(\mathbf{s}).
\end{equation}
The encoder $f_{en}(\cdot)$ in Eq. (\ref{E7}) and the decoder $f_{de}(\cdot)$ in Eq. (\ref{E8}) together constitute the autoencoder model. Suppose the number of elements in $\mathbf{s}$ is $N_{cw}$, and then the compression ratio will be
\begin{equation}
\label{E9}
    \gamma = \frac {N_{cw}}{2 \times N_p \times N_t}.
\end{equation}

\begin{figure}[t]
    \centering
    \resizebox{3in}{!}{%
    \includegraphics*{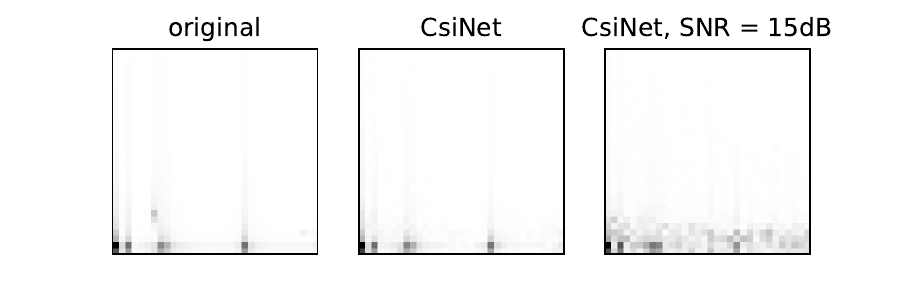} }%
    \caption{\label{fig:add noise}Results when compress ratio is 1/8: (a) original channel; (b) reconstructed channel by CsiNet when noise is free; (c) reconstructed channel by Csinet when SNR = 15dB.}
\end{figure}

The DL-based algorithms \cite{wen2018deep,liu2019exploiting} assume the channel feedback process is perfect. However, the CSI feedback will be impacted by various interference and non-linear effect in practice\cite{martins2008coding}. Similar to \cite{xu2019on}, we take all imperfections as noise (not necessarily Gaussian) and superimpose it on codeword as
\begin{equation}
\label{E10}
    \tilde{\mathbf{s}} = \mathbf{s} + \mathbf{n}.
\end{equation}
In Fig. \ref{fig:add noise}, we compare the gray-scale images of the reconstructed channel at the BS when there exists feedback noise, i.e., when Eq. (\ref{E10}) holds. As shown in Fig. \ref{fig:add noise}(c), when SNR = 15dB, the NMSE increases from -11.55dB to -1.67dB and the gray-scale image of reconstructed CSI exhibits obvious distortion.

	\section{Design of DNNet}\label{sec:deep-model}
In this section, we first describe the details of the proposed DNNet, and then design the joint training mechanism for DNNet and DL-based CSI feedback algorithm.
	\subsection{Structure of DNNet}\label{subsec:detail of the design}

\begin{figure}[t]
    \centering
    \resizebox{2.8in}{!}{%
    \includegraphics*{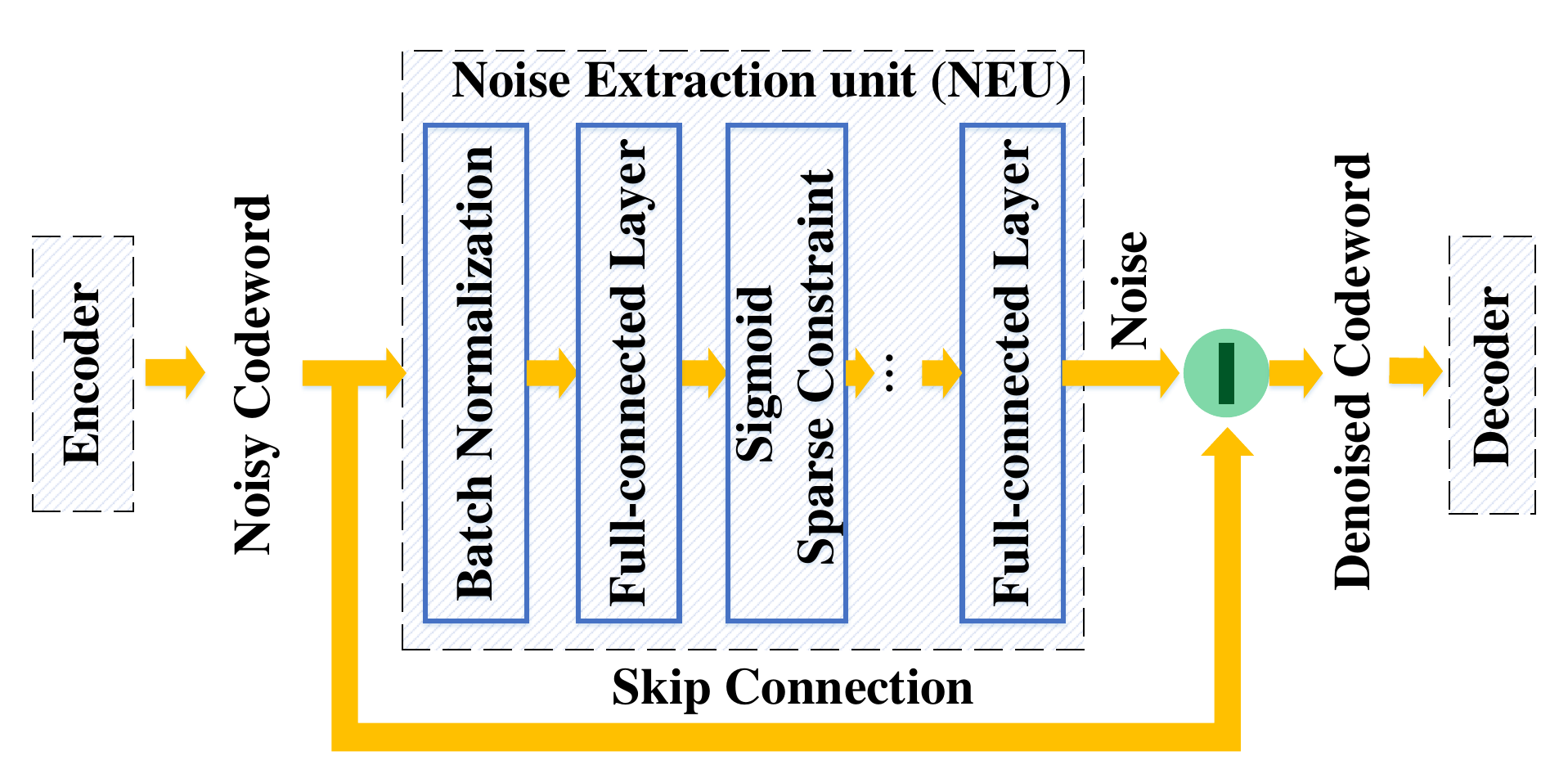} }%
    \caption{\label{fig:DNNet}The structure of DNNet}
\end{figure}

In this paper, we design a specific codeword denoising structure for feedback channel, as in Fig. ~\ref{fig:DNNet}. The basic idea of DNNet is to extract the noise from the codeword by a noise extraction unit (NEU) and then subtract it from the codeword. Details are as follows.

The NEU adopts a fully-connected neural network with $L$ layers, including one input layer, one output layer, and $L-2$ hidden layers. The input codeword is written as
\begin{equation}
\tilde{\mathbf{s}} = [\tilde{s}_1, \tilde{s}_2,\cdots, \tilde{s}_{N_{cw}} ].
\end{equation}

Before inputting the codeword, we first operate the batch normalization (BN) for accelerating deep network training and avoiding output saturation as
\begin{equation}
\tilde{\mathbf{s}}^{'}=\frac{\tilde{\mathbf{s}}-\mathrm{E}\left[\tilde{\mathbf{s}}\right]}{\sqrt{\operatorname{Var}\left[\tilde{\mathbf{s}}\right]}}.
\end{equation}

The output of NEU is the noise that is extracted from codeword, i.e.,
\begin{equation}
\hat{\mathbf{n}}=\boldsymbol{g}_{L}\left(\cdots \boldsymbol{g}_{1}\left(\tilde{\mathbf{s}}^{'}\right)\right).
\end{equation}
The output of the $l$th layer can be written as
\begin{equation}
\boldsymbol{g}_{l}(\mathbf{z})= \left\{\begin{array}{ll}{\mathbf{z}} & {l=1} \\
\bm{\zeta}\left(\boldsymbol{W}_{l} \mathbf{z}+\boldsymbol{b}_{l}\right) & {2 \leq l<L-1} \\
\boldsymbol{W}_{l} \mathbf{z}+\boldsymbol{b}_{l} & {l=L},\end{array}\right.
\end{equation}
where $\bm \zeta(\mathbf{x})=(1+\exp \mathbf{(-x)})^{-1}$ is the sigmoid activation function that adds non-linear transformation. The number of neurons in the hidden layer is 1024, which is generally higher than the dimension of the codeword to capture the noise easily. To avoid overfitting of the high dimensional network, we then add Kullback-Leibler (KL) divergence sparsity constraints to the hidden layer as
\begin{equation}
\label{E11}
\mathrm{KL}(\hat{\beta} \| \beta)=\sum_{j=1}^{|\bm{\hat{\beta}|}} \beta \log \frac{\beta}{\hat{\beta}_{j}}+(1-\beta) \log \frac{(1-\beta)}{1-\hat{\beta}_{j}},
\end{equation}
\begin{equation}
\label{E12}
\bm {\hat{\beta}}=\frac{1}{N} \sum_{i}^{N} {\bm \zeta}\left(\mathbf{W} \mathbf{z}_{i}+\mathbf{b}\right),
\end{equation}
where $|\bm{\hat{\beta}}|$ is the number of neurons, $N$ is the batch size, $\mathbf{z}_{i}$ is the output vector of the $i$th batch from fully-connected layer, $\beta$ is predetermined sparsity factor, and $\hat{\beta}_j$ is the actual sparsity factor of the network.

Moreover, we deliver the codeword to the output of NEU with a skip connection. After noise $\hat{\mathbf{n}}$ is extracted by NEU, the codeword subtracts $\hat{\mathbf{n}}$  to achieve denoising, i.e.,
\begin{equation}
    \hat{\mathbf{s}} = \tilde{\mathbf{s}} - \hat{\mathbf{n}}.
\end{equation}
Then, the denoised codeword is output from the DNNet and is transported to decoder for CSI reconstruction.
    \subsection{Training of DNNet}\label{trainning method of DNNet}

\begin{figure}[t]
    \centering
    \resizebox{3in}{!}{%
    \includegraphics*{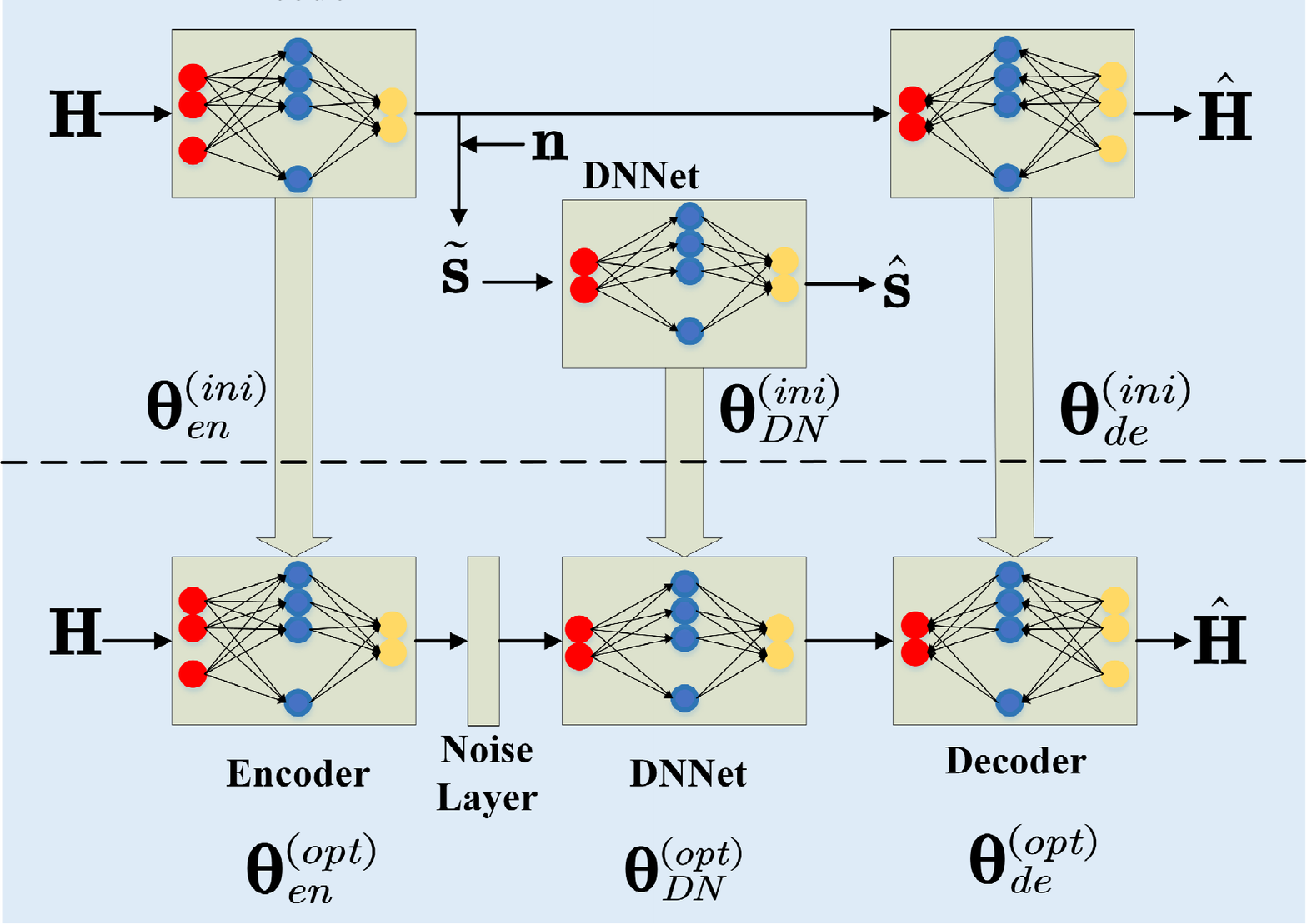} }%
    \caption{\label{fig:Train}(a) Pre-training stage; (b) Joint-training stage}
\end{figure}

The proposed DNNet works together with the existing DL-based channel feedback algorithm \cite{wen2018deep,guo2020convolutional,liu2019exploiting} to enhance feedback performance. Since all existing algorithms use autoencoder architecture, we call the DL-based channel feedback as \textit{autoencoder model} for convenience. Here, we design a novel training mechanism for the autoencoder model and DNNet, which consists of two stages: pre-training stage and joint-training stage, as shown in Fig \ref{fig:Train}.

In the pre-training stage, we train the autoencoder model and DNNet separately to get initial weight coefficients for the next joint-training stage. First, we train the autoencoder model by taking the real and imaginary parts of $\mathbf{H}$ as input data, whose size is $2 \times N_p \times N_t$. The weight coefficients of the autoencoder model are initialized to follow the truncated normal distribution. The loss function is set as MSE
\begin{equation}
\label{E13}
LOSS=\frac{1}{N} \sum_{i=1}^{N}\left\|\hat{\mathbf{H}}_{i}-\mathbf{H}_{i}\right\|_{2}^{2},
\end{equation}
and the adaptive moment estimation (ADAM) algorithm is used to optimize the loss function. Next, we train the DNNet, for which we need to generate the dataset from Eq. (\ref{E7}) and Eq. (\ref{E10}). Specially, in Eq. (\ref{E7}), the encoder is applied to generate the codeword with the aid of well trained autoencoder model while in Eq. (\ref{E10}) the noise is added into the codeword. The MSE loss function and the ADAM optimizer are also utilized, similar to the autoencoder model. After several epochs' training\footnote{\noindent One epoch means training once with all the samples in the training set.}, we can get the initial coefficients $\bm{\theta}_{en}^{(ini)}$, $\bm{\theta}_{de}^{(ini)}$, $\bm{\theta}_{DN}^{(ini)}$, as shown in Fig \ref{fig:Train}(a).

In the joint-training stage, we connect the autoencoder model and DNNet and train them together to get the optimal weight coefficients, as shown in Fig. \ref{fig:Train}(b). The initial weight coefficients of the autoencoder model and DNNet are inherited from the pre-traing stage. In this way, we can get the optimal weight coefficients with fewer training epochs than randomly initializing the weight coefficients at joint-training stage \cite{erhan2010why}. Different from pre-training stage, we need to generate the dataset of DNNet in real time. Hence, we insert a noise layer between the encoder and DNNet to implement Eq. (\ref{E10}), as shown in Fig. \ref{fig:Train}(b). Moreover, an alternate training strategy is employed, where the autoencoder model is frozen when training DNNet while DNNet is frozen when training the autoencoder model. After several iterations, we can get the joint optimal coefficients $\bm{\theta}_{en}^{(opt)}$, $\bm{\theta}_{de}^{(opt)}$, $\bm{\theta}_{DN}^{(opt)}$, as shown in Fig \ref{fig:Train}(b).
\begin{figure*}[t]
\centering     
\subfigure[NMSE v.s. SNR]{\label{NMSE}\includegraphics[height = 57mm,width=76mm]{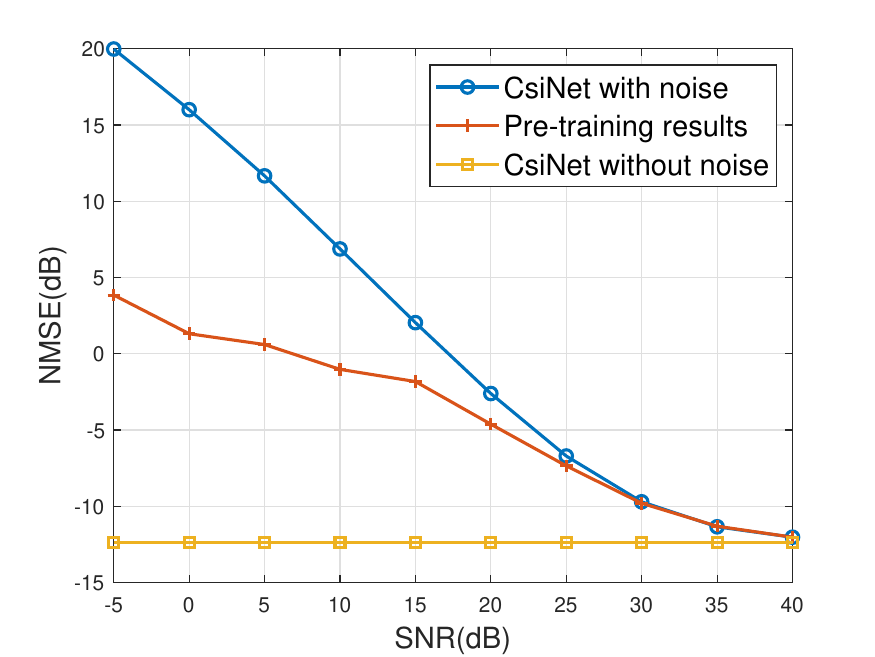}}
\subfigure[$\rho$ v.s. SNR ]{
\label{rho}\includegraphics[height = 57mm,width=76mm]{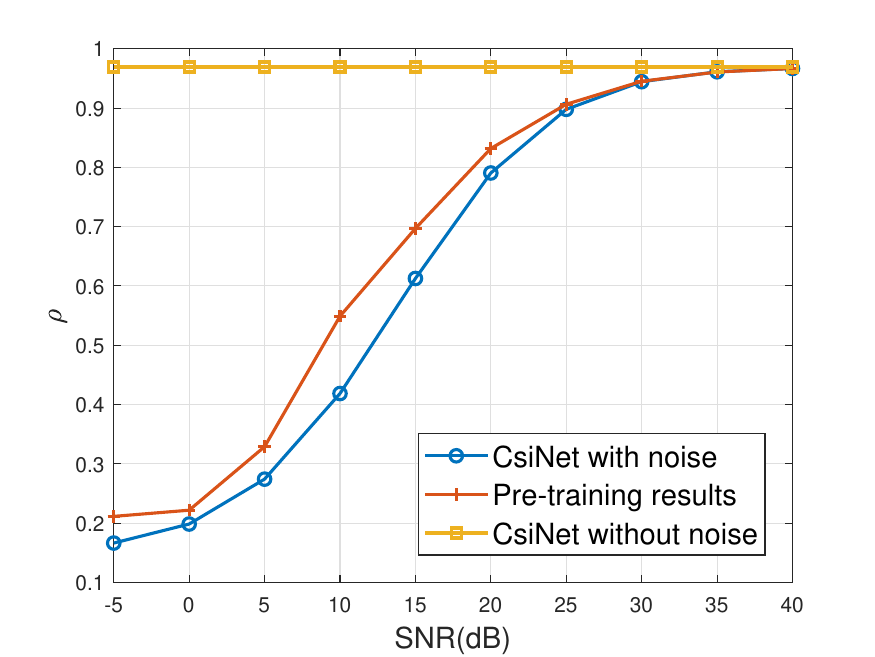}}
\caption{The performance of CsiNet with and without DNNet when $\gamma = 1/8$.}
\label{NMSErho}
\end{figure*}

\subsection{Computational Complexity Analysis}

After the training process, the test process is carried out, where matrix multiplication is the computation-intensive operation and other operation time can be ignored. Assuming that the $l$th layer has $N_l$ neurons, then the computational complexity will be $O(\Sigma_{l=2}^{L}{N_lN_{l-1}})$.
	\section{Simulation Results}\label{sec:simulation}

In our simulation, COST2100 channel model \cite{liu2012the} is used to generate the dataset. Indoor picocellular scenario is deployed, where carrier frequency is 5.3 GHz and subcarrier number $N_c$ is $1024$. The BS is equipped with uniform linear array (ULA) antennas with $N_t = 32$ and is placed at the center of a $20m \times 20m$ square area. The UEs are randomly distributed in this square area. Other parameters follow the default settings in \cite{liu2012the}.

The CSI in spatial-frequency domain is first generated as the COST2100 channel model. Then we transfer it into angular-delay domain and set $N_p = 32$. Moreover, 100,000, 30,000, and 20,000 samples are generated as training set, validation set, and testing set, respectively. The learning rate, batch size are set as 0.001 and 200. The epochs of pre-training and joint-training are 1000 and 500, other parameters in two stages the same. After finishing the training of the autoencoder model, we generate the dataset for DNNet by Eq. (\ref{E7}) and Eq. (\ref{E10}), where noise $\mathbf{n}$ is added into the codeword. Without loss of generality, we assume $\mathbf{n}$ is an i.i.d Gaussian noise vector with $\mathbf{n} \sim \mathcal{N}\left(\mathbf{0}, \sigma_{n}^{2} \mathbf{I}\right)$, and $\sigma_{n}^{2}$ is the noise power. In the second stage, the dataset for DNNet is generated automatically by the noise layer, such that we only need to generate the dataset for the autoencoder model.

Normalized mean-squared error (NMSE) and cosine correlation $\rho$ are used as performance to measure, which are defined as
\begin{equation}
\label{E15}
\mathrm{NMSE} \triangleq \mathrm{E}\left\{\frac{\|\mathbf{H}_{sf}-\hat{\mathbf{H}}_{sf}\|_{2}^{2}}{\|\mathbf{H}_{sf}\|_{2}^{2}}\right\},
\end{equation}
\begin{equation}
\rho \triangleq \mathrm{E}\left\{\frac{1}{{N}_{\mathrm{c}}} \sum_{n=1}^{{N}_{\mathrm{c}}} \frac{|\hat{\mathbf{h}}_{n}^{H} {\mathbf{h}}_{n}|}{\|\hat{\mathbf{h}}_{n}\|_{2}\|{\mathbf{h}}_{n}\|_{2}}\right\},
\end{equation}
where $\mathbf{H}_{sf}$ and $\hat{\mathbf{H}}_{sf}$ are true and reconstructed CSI in spatial-frequency domain, while ${\mathbf{h}}_{n}$ and $\hat{\mathbf{h}}_{n}$ are true and reconstructed CSI of the $n$th subcarrier.

\begin{figure*}[t]
\centering     
\subfigure[NMSE v.s. SNR]{\label{contrnmse}\includegraphics[height = 57mm,width=76mm]{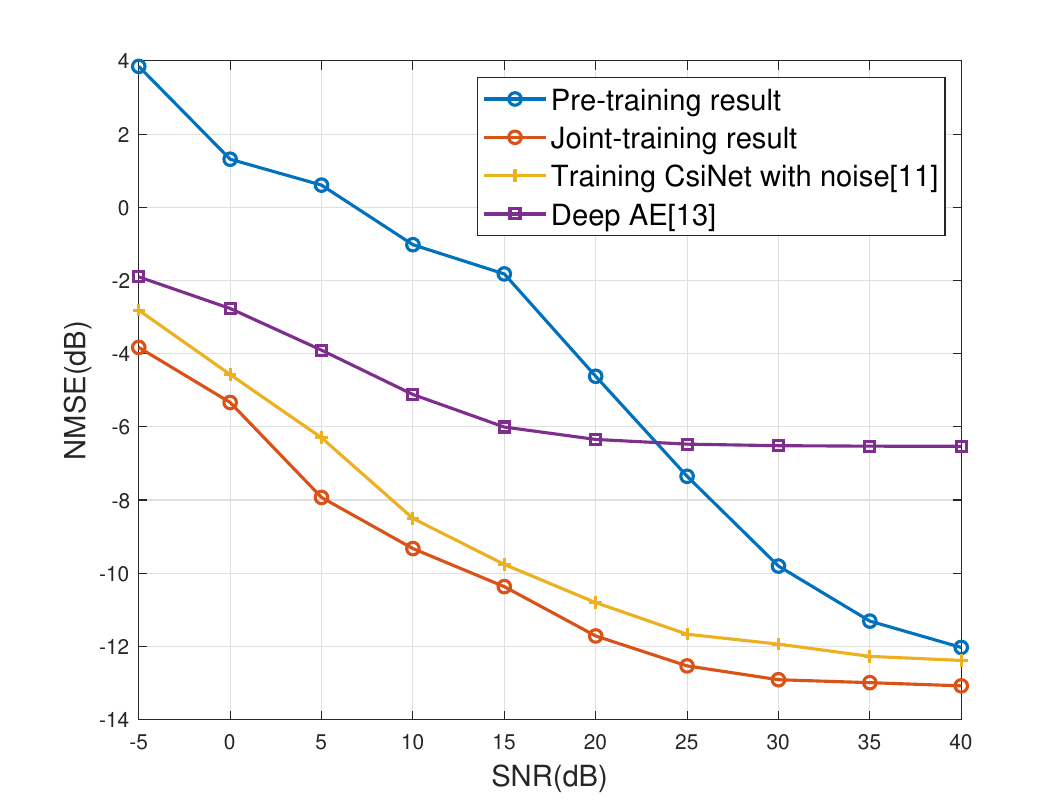}}
\subfigure[$\rho$ v.s. SNR ]{
\label{contrrho}\includegraphics[height = 57mm,width=76mm]{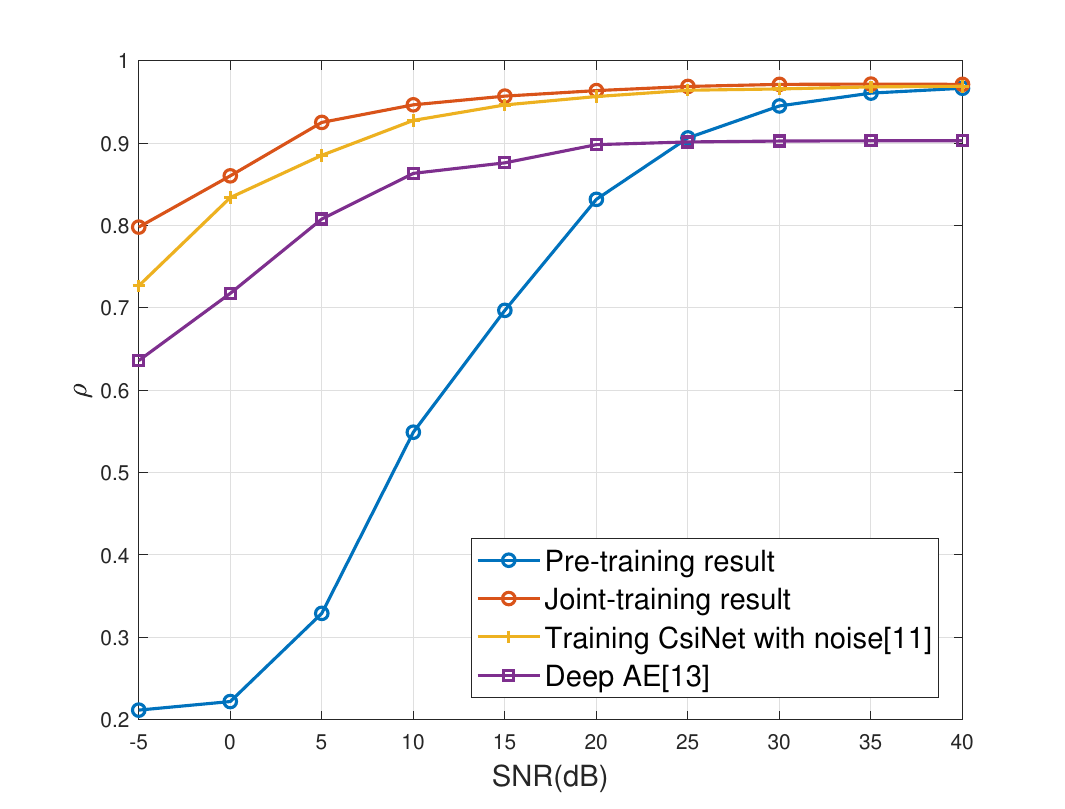}}
\caption{The performance of pre-training, joint-training, Deep AE when $\gamma = 1/8$.}
\label{contrnmserho}
\end{figure*}

\begin{table}[t]
\centering
\caption{NMSE (dB) for different $\gamma$ when codeword is normalized.}
\label{table:normalization}
\begin{footnotesize}
\setlength{\tabcolsep}{3.8mm}{
\begin{tabular}{cl|cccccc}
\hline
\multirow{2}{*}{$\gamma$}  && \multicolumn{2}{c}{CsiNet} & \multicolumn{2}{c}{Normalized CsiNet}  \\
                          &  \multicolumn{1}{c|}{}                         & NMSE (dB)              & $\rho$            & NMSE (dB)              & $\rho$                     \\ \hline\hline
\multirow{1}{*}{1/4}   &                                           & {-17.36}           & 0.99           & {\textbf{-19.17}}         & {\textbf{0.99}}                 \\
                           \hline
\multirow{1}{*}{1/16}     &                                       & {-8.65}           & 0.93           & {\textbf{-9.16}}           & {\textbf{0.94}}                    \\
                          \hline
\multirow{1}{*}{1/32}     &                                       & {-6.24}           & 0.89           & {\textbf{-7.67}}          &{\textbf{0.91}}                 \\
                           \hline
\multirow{1}{*}{1/64}     &                                       & {\textbf{-5.84}}          & {\textbf{0.87}}           & -5.22           & 0.84                   \\
                          \hline
\end{tabular}}
\end{footnotesize}
\end{table}
Without loss of generality, we choose CsiNet as the DL-based feedback algorithms in our simulation\footnote{Actually, the proposed DNNet can work together with any autoencoder models in channel feedback region.}. To quantitatively measure the impact of noise, a normalization layer is inserted after the encoder in CsiNet, i.e.

\begin{equation}
\label{E14}
 \mathbf{s} = \frac{\mathbf{s}}{\|\mathbf{s}\|_{2}}.
\end{equation}
Then, the SNR can be defined as
\begin{equation}
\label{E14}
 SNR = \frac{1}{N_{cw}\sigma_n^{2}},
\end{equation}
where $N_{cw}$ is the length of codeword. We compare the performance when SNR ranges from -5dB to 40dB.

We first compare the NMSE and $\rho$ between the original CsiNet and the normalized CsiNet in perfect feedback channel. As shown in Table \ref{table:normalization}, the better results are presented in bold font and $\gamma$ represents the compression ratio. From Tab. ~\ref{table:normalization}, the normalization will not affect the performance compared to the original CsiNet.

Fig. \ref{NMSE} depicts the NMSE performance of the CsiNet with and without DNNet after pre-training stage, respectively. We use the NMSE of CsiNet in noise-free conditions as a benchmark, which is a horizontal line at -12.38dB. When there exists noise $\mathbf{n}$, the NMSE of CsiNet has a significant degradation compared with noise-free condition. When the SNR increases, the NMSEs of both CsiNet with and without DNNet decrease gradually since the effect of the noise reduces. Nevertheless, the NMSE of CsiNet with DNNet is always lower than that of a single CsiNet. In particular, the NMSE performance of CsiNet with DNNet has an improvement about 5-10 dB compared with a single CsiNet at low SNR.

Fig. \ref{rho} shows the performance of $\rho$ at different SNR. When there exists noise $\mathbf{n}$, the performance of both CsiNet with and without DNNet gradually approaches the that in the noise-free condition as SNR increases. Nevertheless, the performance of the single CsiNet is always lower than CiNet with pre-training DNNet. When SNR is over 35dB, the performance of both CsiNet with and without DNNet approaches the benchmark.

The joint-training stage can further improve the performance of CsiNet with DNNet. As shown in Fig. \ref{contrnmse}, we compare the NMSE performance of the CsiNet with DNNet after pre-training stage, the CsiNet with DNNet after joint-training stage, CsiNet trained with noise \cite{guo2020convolutional} and DeepAE \cite{jang2019deep}. When training the autoencoder model, authors in \cite{guo2020convolutional} take the feedback noise into account. The DeepAE employs a three-layer neural network as encoder and a symmetrical three-layer neural network as decoder. Similarly, \cite{jang2019deep} take the feedback noise into account in the training phase. With the increase of SNR, the NMSE of the algorithms gradually decreases. After joint-training stage, CsiNet with DNNet significantly outperforms that after pre-training stage at all SNRs, which indicates that the performance of the network against noise is enhanced by DNNet after joint-training stage. Besides, at all SNRs, CsiNet with jointly trained DNNet outperforms the other algorithms.


The trend of $\rho$ versus SNR is similar to NMSE. As shown in Fig. \ref{contrrho}, the $\rho$ of the algorithms gradually increases with SNR. Moreover, at all SNRs, CsiNet with jointly trained DNNet exhibits the best performance among the algorithms.


The NMSE and $\rho$ of the three algorithms in Fig. \ref{contrnmserho} imply that the proposed DNNet with two-stage training can combat noise and get the best performance among the existing algorithms that consider the imperfect feedback channel.

	\section{Conclusion}\label{sec:conclusion}
In this article, we design a DL-based denoise network, named DNNet, to improve the performance and robustness of channel feedback. We propose a training mechanism for DNNet by jointly training the existing DL-based feedback algorithm and DNNet. Numerical results have demonstrated that the proposed DNNet with two-stage training mechanism can perform better than the existing algorithms that consider the imperfect feedback channel. Future work includes using more advanced deep learning algorithms to achieve better denoising performance.



\end{document}